\documentclass[english,aps,prb,twocolumn,amsmath,amssymb,showpacs,superscriptaddress,notitlepage]{revtex4-1}
\usepackage[T1]{fontenc}
\usepackage[latin9]{inputenc}
\usepackage{amsmath}
\usepackage{amssymb}
\usepackage{graphicx}
\usepackage{esint}

\makeatletter
\usepackage{amsmath,amssymb,amsfonts,bm}
\usepackage{graphicx}
\usepackage{epstopdf}
\usepackage{dcolumn}
\usepackage{mathrsfs}
\usepackage{bbold}
\usepackage{dsfont}
\usepackage{dcolumn}

\usepackage[colorlinks=true,linkcolor=blue,citecolor=blue, urlcolor=blue,bookmarks=false]{hyperref}

\makeatother

\usepackage{babel}
\begin{document}
\title{On the half-quantized Hall conductance of massive surface electrons
in magnetic topological insulator films}
\author{Rui Chen}
\address{Department of Physics, The University of Hong Kong, Pokfulam Road,
Hong Kong 999077, China}
\affiliation{Department of Physics, Hubei University, Wuhan 430062, China}

\author{Shun-Qing Shen}
\email{sshen@hku.hk}

\affiliation{Department of Physics, The University of Hong Kong, Pokfulam Road,
Hong Kong 999077, China}
\date{\today}
\begin{abstract}
In topological insulators, massive surface bands resulting from local
symmetry breaking are believed to exhibit a half-quantized Hall conductance.
However, such scenarios are obviously inconsistent with the Thouless-Kohmoto-Nightingale-Nijs
theorem, which states that a single band in a lattice with a finite
Brillouin zone can only have an integer-quantized Hall conductance.
To explore this, we investigate the band structures of a lattice model
describing the magnetic topological insulator film that supports the
axion insulator, Chern insulator, and semi-magnetic topological insulator
phases. We reveal that the gapped and gapless surface bands in the three
phases are characterized by an integer-quantized Hall conductance
and a half-quantized Hall conductance, respectively. This result
is distinct from the previous consensus that the gapped surface band
is responsible for the half-quantized Hall conductance and the gapless
band should exhibit zero Hall response. We propose an effective model
to describe the three phases and show that the low-energy dispersion
of the surface bands inherits from the surface Dirac fermions. The
gapped surface band manifests a nearly half-quantized Hall conductance
at low energy near the center of Brillouin zone, but is compensated by another nearly half-quantized
Hall conductance at high energy near the boundary of Brillouin zone because a single band can only have
an integer-quantized Hall conductance. The gapless state hosts a zero
Hall conductance at low energy but is compensated by another half-quantized
Hall conductance at high energy, and thus the half-quantized Hall
conductance can only originate from the gapless band. Moreover, we
calculate the layer-resolved Hall conductance of the system. The conclusion
suggests that the individual gapped surface band alone does not support
the half-quantized surface Hall effect in a lattice model.
\end{abstract}
\maketitle
\section{Introduction}
The half-quantized Hall effect occurs on the
topological insulator surface once the surface Dirac fermion acquires
a mass due to local symmetry breaking\,\citep{Qi08PRB,Nomura2011PRL,Qi2011RMP,Haldane1988PRL,Fu07PRL,Niemi83PRL,Jackiw84PRD,Semenoff84PRL}
for three reasons: (i) The bulk value of the axion angle $\theta=\pi$
in topological insulators allows either gapless or gapped surface
states\,\citep{Qi08PRB,Qi2011RMP,Mong10prb,Wilczek87PRL,Essin09prl,Hasan2010RMP}.
In the gapped case, there must be a half-quantized Hall effect; (ii)
The half-quantized Hall effect can be captured by calculating the
layer-resolved Hall conductance\,\citep{Mong10prb,Wang15prbrc,Varnava18prb,Essin09prl,FuB2021PRR};
and (iii) The gapped surface state could be described by the massive
Dirac equation, which hosts a half-quantized Hall conductance\,\citep{Chu11prb,liu2021dissipative,Sun20prbrc,Shen2017TI,Shan2010NJP,Lu13prl-QAH}.
Therefore, it is widely believed that the half-quantized Hall conductance
originates from the gapped surface state in magnetic topological insulators\,\citep{mogi2021experimental,Gao2021Nature,Mogi17sa,Hao19prx}.
However, there are still concerns for the following reasons: (i) The
axion term and (ii) layer-resolved Hall conductance do not guarantee
that the half-quantized Hall effect originates from the gapped surface
state; (iii) The half quantization in the massive Dirac equation contradicts
the common belief that a single band on a two-dimensional finite Brillouin
zone can only have an integer-quantized Hall conductance. Moreover,
our recent works focus on ``parity anomalous semimetals'', in which
massive and massless Dirac fermions coexist\,\citep{Zou22PRB,ZouJY2022arXiv,fu2022arXiv}.
The half-quantized Hall conductance in the parity anomalous semimetal
is attributed to the massless Dirac fermions, rather than the previous
consensus that it should be attributed to the massive Dirac fermions.
Therefore, how the half quantization is manifested in magnetic topological
insulator films is still an open question.

The half-quantized Hall effect is proposed to be realized on magnetically
doped topological insulators\,\citep{Yu2010Science,Chang2013Science,Chang2015NatMat,Mogi15apl,Mogi17nm,Mogi17sa,Xiao18prl,Tokura19nrp,Zhao13srep,Fijalkowski2021PRB,ChenCZ21PRL,Liu20prb,Varnava18prb,mogi2021experimental,zhou2022axion,Zou22PRB,Liu2021Adv}
and intrinsic antiferromagnetic topological insulators\,\citep{Deng20sci,Zhang20prl,Sun20prbrc,
Shi19prb,Rienks19nat,Otrokov19prl,
Lin2022PRB,li2021nonlocalarXiv,Li19sa,
Li19prx,Li19prbrc,Li19prb,Lee19prr,
Chen19nc,Chen19prx,Hao19prx,Ding20prbrc,
Mong10prb,Otrokov19nat,ChenR21PRB,
Gu2021Nc,Liu20nm,Gao2021Nature,ChenR2022NSR,lei2023kerr,mei2023electrically}.
According to the electronic band structures, the magnetic topological
insulator can be divided into three different phases: the semi-magnetic
topological insulator phase\,\citep{mogi2021experimental,zhou2022axion,Zou22PRB}
consists of a gapped and a gapless Dirac cone on its top and bottom
surfaces {[}Fig.\,\ref{Fig_illustration}(b){]}; the axion\,\citep{Liu20nm,Mogi17nm,Mogi17sa,Xiao18prl,ChenR2022NSR,Gao2021Nature}
and Chern\,\citep{Chu11prb,ChenR21PRB,Chang2013Science,Yu2010Science,Gu2021Nc,Liu20nm}
insulator phases host two gapped Dirac cones on the top and bottom
surfaces {[}Figs.\,\ref{Fig_illustration}(a) and \ref{Fig_illustration}(c){]}.
Specifically, the gapped Dirac cones on the axion (Chern) insulator
phase were believed to be characterized by a half-quantized Hall conductance
with the opposite (same) sign on the opposite surface, and they combine
to yield a zero (quantized) Hall conductance. However, such scenarios are obviously inconsistent with the Thouless-Kohmoto-Nightingale-Nijs
theorem, which states that a single band in a lattice with a finite
Brillouin zone can only have an integer-quantized Hall conductance. In the semi-magnetic
topological insulator phase, the magnetic gap is characterized by
a half-quantized Hall conductance contributed by the gapped surface.
The gapless surface is believed to exhibit no effect on the total
Hall conductance. In the past years, the three phases have received
enormous attention. The axion insulator possesses a unique electromagnetic
response from the massive Dirac fermions, which leads to the quantized
topological magnetoelectric effect\,\citep{Wang15prbrc,Morimoto15prb,Qi09sci,Maciejko10prl,Tse10prl,Yu19prb}.
The semi-magnetic topological insulator provides a condensed-matter
realization of the parity anomaly and opens the door for experimental
realization of the single Dirac fermion\,\citep{mogi2021experimental,zhou2022axion,Zou22PRB}.
The realization of the Chern insulator phase can lead to the development
of novel spintronics devices\,\citep{Yu2010Science,Chang2013Science}.

In this work, we systematically study the spectrum, probability distribution,
and the corresponding Berry curvature distribution in three phases
in magnetic topological insulator films, including the axion insulator,
Chern insulator, and semi-magnetic topological insulator phases. We
explore how the band structures manifest the half-quantized Hall conductance
in a lattice model. In the axion and Chern insulator phases, each
gapped surface state contributes a nearly half-quantized Hall conductance
at low energy near the center of Brillouin zone, but must be compensated by another nearly half-quantized
Hall conductance at high energy near the boundary of Brillouin zone, because a single band can only have
an integer-quantized Hall conductance in a lattice model according
to Thouless-Khomoto-Nightingale-Nijs theorem\,\citep{Thouless1982PRL}.
In the semi-magnetic topological insulator phase, the gapless state
hosts a zero Hall conductance at low energy but must be compensated
by another half-quantized Hall conductance at high energy, because
the half-quantized Hall conductance can only originate from the gapless
state. Moreover, we propose an effective four-band model to depict
the three phases. We confirm that the low-energy surface states in
these phases inherit from the Dirac fermions. The states at low and
high energies always couple in energy scale and they cannot be treated
separately. Due to the coupling effect, the exact half quantization
is only revealed in the gapless surface state in the semi-magnetic
topological insulator phase, but it is not found in the gapped surface state
in the axion and Chern insulator phases from the band calculation.

Moreover, we adopt the layer-resolved Hall conductance to study the
three phases. The three phases are characterized by distinct patterns
of half-quantized surface Hall effect. The results indicate that the
half-quantized surface Hall effect is contributed by all the occupied
bands, rather than the individual gapped surface bands.

\section{Model}

To carry out numerical studies on the magnetic topological insulator,
we consider a tight-binding Hamiltonian on a cubic lattice for an
isotropic three-dimensional topological insulator~\citep{Chu11prb,Shan2010NJP,LuHZ2010PRB}
\begin{equation}
\mathcal{H}=\sum_{i}c_{i}^{\dagger}\mathcal{M}_{0}c_{i}+\sum_{i,\alpha=x,y,z}\left(c_{i}^{\dagger}\mathcal{T}_{\alpha}c_{i+\alpha}+c_{i+\alpha}^{\dagger}\mathcal{T}_{\alpha}^{\dagger}c_{i}\right),\label{Eq_1}
\end{equation}
where $\mathcal{T}_{\alpha}=B\sigma_{z}\tau_{0}-i\frac{A}{2}\sigma_{x}\tau_{\alpha}$
and $\mathcal{M}_{0}=(M-6B)\sigma_{z}\tau_{0}+\Delta\left(z\right)\sigma_{0}\tau_{z}$
with the lattice space is taken to be unity. Near the $\boldsymbol{k}=0$
point in the momentum space (i.e., the low-energy regime), this model
is reduced to a Dirac-like model in the absence of the magnetic effect.
$\sigma$ and $\tau$ are Pauli matrices. The magnetic effect is represented
by layer-dependent Zeeman splitting $\Delta\left(z\right)$. We take
$\Delta\left(z\right)=\Delta_{b}$ for the bottom surface with $z=1,2$,
$\Delta\left(z\right)=\Delta_{t}$ for the top surface with $z=n_{z}-1,n_{z},$
and $\Delta\left(z\right)=0$ elsewhere. The axion and Chern insulator
phases are characterized by $\Delta_{t}\Delta_{b}<0$ and $\Delta_{t}\Delta_{b}>0$,
respectively. The semi-magnetic topological insulator phase is characterized
by either $\Delta_{b}\neq0$ and $\Delta_{t}=0$ or $\Delta_{b}=0$
and $\Delta_{t}\neq0$. In the subsequent calculations, we fix the
parameters as $A=0.5$, $B=0.25$, and $M=0.4$.

\begin{figure}
\includegraphics[width=8.5cm]{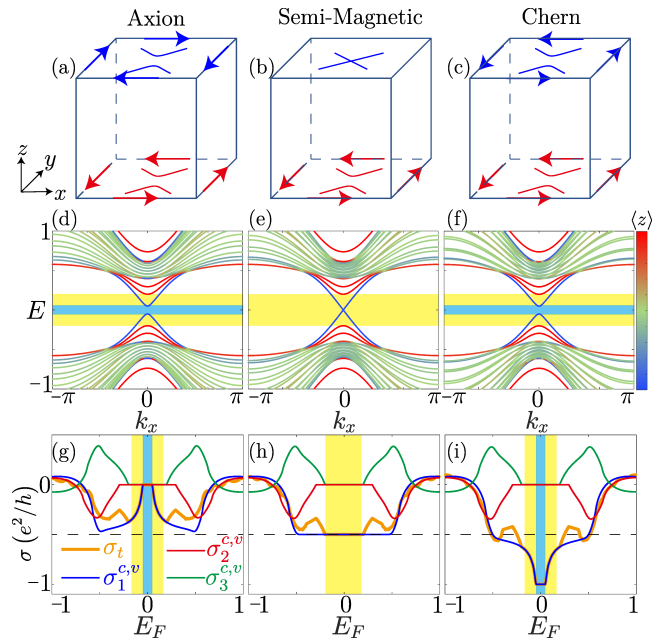} \caption{\label{Fig_illustration}(a) Schematic illustration of the axion insulator
phase. The arrows indicate the propagating direction of the surface
chiral currents. Blue and red correspond to the top and bottom surfaces,
respectively. (d) Energy spectra of the axion insulator phase. Here
the color scheme of the bands indicates the wave function distribution.
(g) Numerically calculated Hall conductance $\sigma_{1}^{c,v}$ (blue),
$\sigma_{2}^{c,v}$ (red), $\sigma_{3}^{c,v}$ (green), and $\sigma_{t}$
(orange) as functions of $E_{F}$. Here, $\sigma_{t}$ depicts the
conductance contributed from all the bands, $\sigma_{1}^{c,v}$ depict
the conductance contributed from the lowest conduction and highest
valence bands, $\sigma_{2}^{c,v}$ depict the conductance contributed
from the second lowest conduction and second highest valence bands,
and so on. The dashed black line corresponds to $\sigma=-e^{2}/2h.$
In (d) and (g), the yellow and light blue regions correspond to the
magnetic gap on the top and bottom surfaces, respectively. (b, e,
h) and (c, f, i) are the same as (a, d, g), except that they depict
the semi-magnetic topological insulator phase and the Chern insulator
phase, respectively. We take $\Delta_{t}=0.05$ for the axion insulator
phase, $\Delta_{t}=0$ for the semi-magnetic topological insulator
phase, and $\Delta_{t}=-0.05$ for the Chern insulator phase. The
Zeeman splitting term for the bottom surface is $\Delta_{b}=-0.2$
for all three phases. The film thickness is taken as $n_{z}=10$.}
\end{figure}

When magnetization is introduced to a certain surface, it is believed
that the gapless Dirac cone will open an energy gap characterized
by a half-quantized Hall conductance, with its sign depending on the
magnetization direction. For both Chern and axion insulator phases,
the Zeeman effect is introduced to the top and bottom surfaces. Therefore
the surface states open energy gaps at the $\Gamma$ point on both
the top and bottom surfaces {[}Figs.\,\ref{Fig_illustration}(a)
and\,\ref{Fig_illustration}(c){]}. When the two surfaces have antiparallel
magnetization alignment\,{[}Fig.\,\ref{Fig_illustration}(a){]},
the system is characterized by half-quantized Hall conductance with
the opposite sign on the opposite surface, leading to the emergence
of the axion insulator phase. Because the top and bottom surfaces
have parallel magnetization alignment in the Chern insulator phase,
the system is characterized by the half-quantized Hall conductance
with the same sign\,{[}Fig.\,\ref{Fig_illustration}(c){]}, and
they combine to yield a quantized Hall conductance. The semi-magnetic
topological insulator phase is characterized by a half-quantized Hall
conductance, contributed by the gapped Dirac cone on the magnetic
bottom surface\,{[}Fig.\,\ref{Fig_illustration}(b){]}.

However, the above scenarios seem to contradict the common belief
that a single band in a lattice model can only host an integer-quantized
Hall conductance in units of $e^{2}/h.$ Next, we study the spectrum
and the corresponding probability distribution of the three phases
in the lattice model in Eq.\,(\ref{Eq_1}), and explore how the band
structures reconcile the contradiction and how the half quantization
is manifested.

\section{Spectrum and Hall conductance}

The second and third rows in Fig.\,\ref{Fig_illustration} show the
numerically calculated energy spectra and the corresponding Hall conductance
for the three phases, respectively. We take $\Delta_{b}=0.2$ and
$\Delta_{t}=\mp0.05$ for the axion and Chern insulator phases and
$\Delta_{b}=0.2$ and $\Delta_{t}=0$ for the semi-magnetic topological
insulator phase, respectively. The results are as expected. In the
axion insulator phase, the spectrum opens a gap {[}Fig.\,\ref{Fig_illustration}(d){]}
characterized by a zero Hall conductance {[}the orange line in Fig.\,\ref{Fig_illustration}(g){]}.
The spectrum of the Chern insulator phase {[}Fig.\,\ref{Fig_illustration}(f){]}
is the same as the axion insulator phase, except that the gap is characterized
by a quantized Hall conductance {[}the orange line in Fig.\,\ref{Fig_illustration}(i){]}.
The spectrum of the semi-magnetic topological insulator phase is gapless
{[}Fig.\,\ref{Fig_illustration}(e){]}, and the system is characterized
by a half-quantized Hall conductance when the Fermi energy is located
at the magnetic gap of the bottom surface {[}the orange line in Fig.\,\ref{Fig_illustration}(h){]}.
Here, the conductance for the $m$-band is calculated by~\citep{Shen2017TI}
\begin{align}
\sigma_{m}\left(E_{F}\right) & =-i\hbar e^{2}\sum_{n\neq m}\intop\frac{d^{2}k}{\left(2\text{\ensuremath{\pi}}\right)^{2}}\frac{\left\langle m\right|v_{x}\left|n\right\rangle \left\langle n\right|v_{y}\left|m\right\rangle }{E_{m}-E_{n}}\nonumber \\
 & \times\frac{f\left(E_{F}-E_{n}\right)-f\left(E_{F}-E_{m}\right)}{E_{m}-E_{n}+i\eta},
\end{align}
wth $\left|m\right\rangle $ is the eigensate of energy $E_{m}$ for
the quasi-2D system confined along the $z$ direction. $v_{x}$ and
$v_{y}$ are the velocity operators. $f\left(x\right)$ is the Fermi
distribution and $\eta$ is an infinitesimal quantity. The total conductance
contributed from all the bands has the form $\sigma_{t}\left(E_{F}\right)=\sum_{m}\sigma_{m}\left(E_{F}\right).$

\begin{figure}
\includegraphics[width=8.5cm]{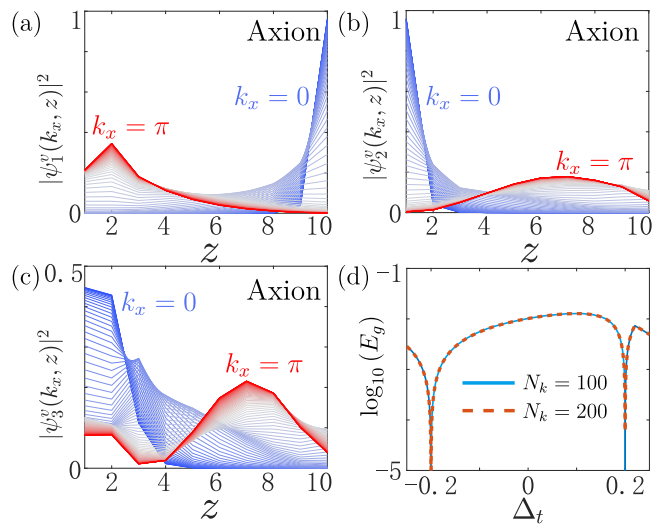} \caption{\label{Fig_wavefunction}(a) Probability distribution $\left|\psi_{1}^{v}\left(k_{x},z\right)\right|^{2}$
as a function of layer index $z$ for the highest valence band of
the axion insulator phase for different $k_{x}.$ (b) and (c) are
the same as (a), except that they depict the second and third highest
valence bands. The probability distributions of the semi-magnetic
topological insulator phase and the Chern insulator phase are similar
to that of the axion insulator phase. (d) The logarithm of the energy
gap between the first and second valence bands as a function of $\Delta_{t}$.
Each point is obtained by searching the minimum energy difference
between the first and second valence bands by scanning in the two-dimensional
Brillouin zone. The $k$ points used in the calculation is $N_{k}\times N_{k}$.
The Zeeman splitting for the bottom surface is taken as $\Delta_{b}=-0.2$.
The film thickness is $n_{z}=10$. In (a-c), we take $\Delta_{t}=0.05$.}
\end{figure}

When the Fermi energy is located at the magnetic surface gaps {[}the
yellow region in Figs.\,\ref{Fig_illustration}(d-f){]}, the topology
nature of the three phases is dominated by the lowest conduction and
highest valence bands {[}see the blue and orange lines in Fig.\,\ref{Fig_illustration}(g-i){]}.
The higher conduction and lower valence bands exhibit no contribution
to the total Hall conductance {[}the red and green lines in Fig.\,\ref{Fig_illustration}(g-i){]}.

Figures\ \ref{Fig_wavefunction}(a-c) show the probability distributions
$\lvert\psi_{i=1,2,3}^{v}\left(k_{x},z\right)\rvert^{2}$ of the valence
bands as functions of $z$ for different $k_{x}$ in the axion insulator
phase, where $\psi_{i}^{v}$ is the wave function distribution of
the $i$-th highest valence band. It is noted that the probability
distributions of the semi-magnetic topological insulator phase and
Chern insulator phase are similar to that of the axion insulator phase
{[}as shown in Figs.\ref{Fig_illustration}(d-f){]}. The probability
distribution of the highest valence band $\lvert\psi_{1}^{v}\left(k_{x},z\right)\rvert^{2}$
is mainly distributed at the top surface for the low-energy states
near $k_{x}=0$ {[}the blue lines in Fig.\,\ref{Fig_wavefunction}(a){]},
and is mainly distributed at the bottom surface for the high-energy
states near $k_{x}=\pi$ {[}the red lines in Fig.\,\ref{Fig_wavefunction}(a){]}.

The second and third highest valence bands exhibit no contribution
to the total Hall conductance when the Fermi energy resides in the
magnetic gap. Their probability distributions $\lvert\psi_{2,3}^{v}\left(k_{x},z\right)\rvert^{2}$
are mainly distributed at the bottom surfaces for the low-energy state
near $k_{x}=0$ {[}the blue lines in Figs.\,\ref{Fig_wavefunction}(b)
and \ref{Fig_wavefunction}(c){]}, and are mainly distributed at the
bulk for the high-energy state near $k_{x}=\pi$ {[}the red lines
in Figs.\,\ref{Fig_wavefunction}(b) and \ref{Fig_wavefunction}(c){]}.
Moreover, to confirm that the first and second valence bands are well
separated in energy scale, we plot $\log_{10}\left(E_{g}\right)$
as a function of $\Delta_{t}$ in Fig.\,\ref{Fig_wavefunction}(d),
where $E_{g}$ is the minimum energy difference between the first
and second valence bands by scanning the whole two-dimensional Brillouin
zone. $E_{g}$ converges with the increasing density of the $k$ points.
The results indicate that the first and second valence bands are well
separated in energy scale, except at two special points with $\Delta_{t}=\pm\Delta_{b}$.

The above scenario indicates that the highest valence band, that dominates
the topology of the three phases when Fermi energy resides in the
magnetic gap, is localized at one surface at low energy near the Dirac
point and at the other surface at high energy near the Brillouin boundary.
In contrast to the previous consensus\,\citep{Li19sa,Wang15prbrc,Morimoto15prb},
the topology natures of the three phases are dominated by the two
surface states near the $\Gamma$ point.

\section{Effective model}

\begin{figure}[t]
\includegraphics[width=7cm]{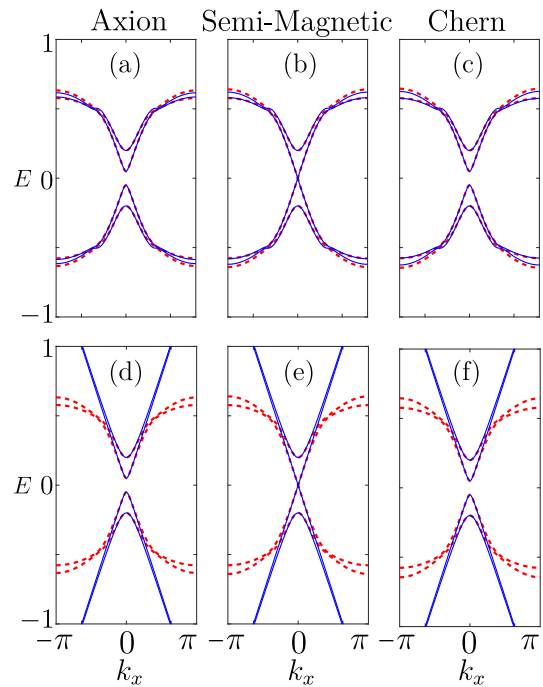}

\caption{\label{Fig_effectivemodel_spectrum}(a, d) Energy spectrum of the
axion insulator phase with $\Delta_{t}=0.05$ and $\Delta_{b}=-0.2$.
The red dashed lines are obtained by numerical calculations (where
only the two lowest conductance and two highest valence bands are
illustrated). The solid lines in (a) and (d) are obtained from the
effective Hamiltonians in Eqs.\ (\ref{eq:Effective_model_1}) and
(\ref{eq:Effective_model_2}), respectively. (b, e) and (c, f) are
the same as (a, d), except that they depict (b, e) the semi-magnetic
topological insulator phase with $\Delta_{t}=0$ and (c, f) the Chern
insulator phase with $\Delta_{t}=0.05$, respectively.}
\end{figure}

The lowest four bands of the magnetic topological insulator thin film
can be effectively described by a $4\times4$ Hamiltonian in the two-dimensional
Brillouin zone\,\citep{LuHZ2010PRB,Zou22PRB,ZouJY2022arXiv,Shan2010NJP}:

\begin{equation}
H=\left(\begin{array}{cc}
h\left(\mathbf{{k}}\right)+V_{t}\tau_{z} & j\left[m_{0}\left(\mathbf{{k}}\right)/T\right]m_{0}\left(\mathbf{{k}}\right)\\
j\left[m_{0}\left(\mathbf{{k}}\right)/T\right]m_{0}\left(\mathbf{{k}}\right) & -h\left(\mathbf{{k}}\right)+V_{b}\tau_{z}
\end{array}\right),\label{eq:Effective_model_1}
\end{equation}
where $m_{0}\left(\mathbf{{k}}\right)=M-4B\left(\sin^{2}k_{x}/2+\sin^{2}k_{y}/2\right)$,
$V_{t/b}\left(k\right)=\Delta_{t/b}j\left[-m_{0}\left(\mathbf{{k}}\right)/T\right]+\Delta'_{t/b}j\left[m_{0}\left(\mathbf{{k}}\right)/T\right]$,
and $h\left(\mathbf{{k}}\right)=A\left(\tau_{y}\sin k_{x}-\tau_{x}\sin k_{y}\right)$
describes the massless Dirac fermion. The Fermi-Dirac-distribution
like factor $j(x)=\left[\exp\left(x\right)+1\right]^{-1}$ describes
the process that the surface states merge into the bulk states. $\Delta'_{t/b}/\Delta_{t/b}\simeq n_{z}^{\textrm{Mag}}/n_{z}\ll1$
where $n_{z}^{\textrm{Mag}}$ is the thickness of the magnetically
doped film. For thick films with a large $n_{z}$, we have $\Delta'_{t/b}\rightarrow0.$
The coefficient $T^{*}=0.05$ is a model-specific parameter. The first
row in Fig.\ \ref{Fig_effectivemodel_spectrum} compares the spectra
obtained from the numerical tight-binding calculations and the analytical
effective models in Eq.\ (\ref{eq:Effective_model_1}). The effective
Hamiltonian captures the band structures of the three phases in the
magnetic topological insulator. Moreover, in the following content,
we will show that this effective Hamiltonian also captures the band
topologies of the three topological phases.

The Brillouin zone is divided into two regimes by the sign of $m_{0}(\mathbf{k})$.
In the low-energy regime near the center of Brillouin zone, i.e., $k^{2}<k_{c}^{2}$, we have $j\left[m_{0}\left(\mathbf{{k}}\right)/T\right]\rightarrow0$
and $j\left[-m_{0}\left(\mathbf{{k}}\right)/T\right]\rightarrow1$,
where the value of $k_{c}$ is given by $m_{0}(\mathbf{k}_{c})=0$.
The effective Hamiltonian in Eq.\ (\ref{eq:Effective_model_1}) reduces
to
\begin{equation}
H=\left(\begin{array}{cc}
h'\left(\mathbf{{k}}\right)+\Delta_{t}\tau_{z} & 0\\
0 & -h'\left(\mathbf{{k}}\right)+\Delta_{b}\tau_{z}
\end{array}\right),\label{eq:Effective_model_2}
\end{equation}
with $h'\left(\mathbf{{k}}\right)=A\left(\tau_{y}k_{x}-\tau_{x}k_{y}\right)$.
The low-energy effective model describes two decoupled Dirac fermions
with the Dirac mass determined by $V_{t}$ and $V_{b}$, respectively.
The effective low-energy effective Hamiltonian reproduces the key
features of the band structures, that the spectra are dominated by
the two decoupled Dirac fermions near the $\Gamma$ point {[}the second
row in Fig.\ \ref{Fig_effectivemodel_spectrum}{]}.

In the high-energy regime near the boundary of Brillouin zone, i.e., $k^{2}>k_{c}^{2}$, we have $j\left[m_{0}\left(\mathbf{{k}}\right)/T\right]\rightarrow1$
and $j\left[-m_{0}\left(\mathbf{{k}}\right)/T\right]\rightarrow0.$
The effective Hamiltonian in Eq.\ (\ref{eq:Effective_model_1}) reduces
to
\begin{equation}
H=\left(\begin{array}{cc}
h\left(\mathbf{{k}}\right)+\Delta'_{t}\tau_{z} & m_{0}\left(\mathbf{{k}}\right)\\
m_{0}\left(\mathbf{{k}}\right) & -h\left(\mathbf{{k}}\right)+\Delta'_{b}\tau_{z}
\end{array}\right).\label{eq:Effective_model_3}
\end{equation}
 The top and bottom surface states are coupled via the term $m_{0}(\mathbf{k})$
in the high-energy regime.

\begin{figure*}
\includegraphics[width=17cm]{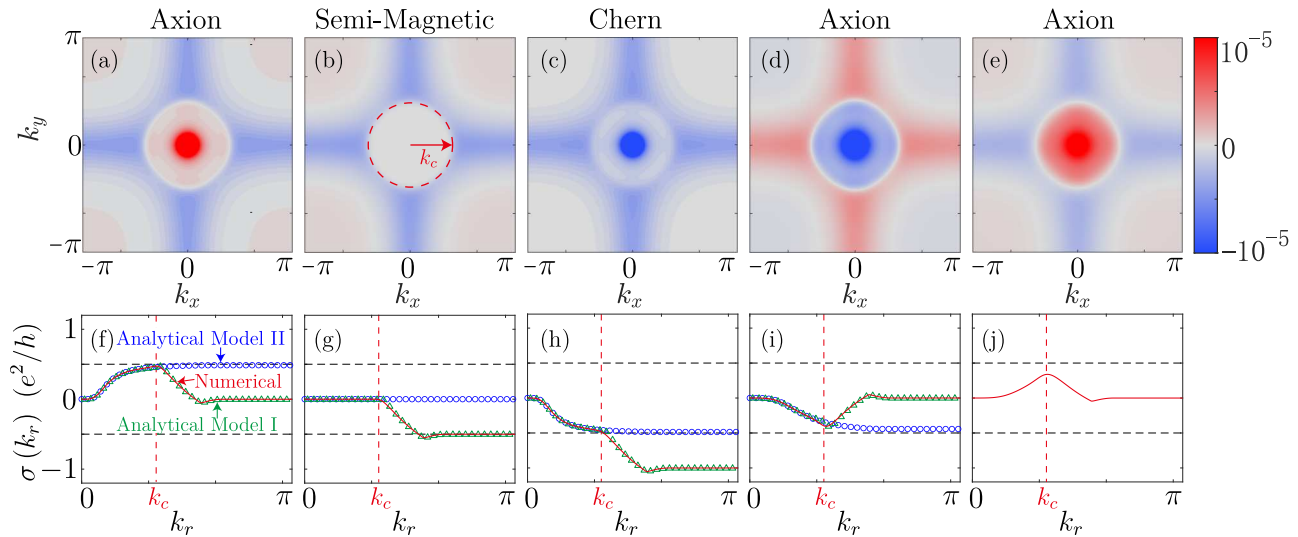}

\caption{\label{Fig_BerryCurvature}(a) $\Omega_{1}^{v}(k_{x},k_{y})$ as a
function of $k_{x}$ and $k_{y}$, where $\Omega_{1}^{v}(k_{x},k_{y})$
corresponds to the Berry curvature of the highest valence band in
the axion insulator phase. (f) $\sigma\left(k_{r}\right)$ as a function
of $k_{r}$. (b, g) and (c, h) are the same as (a, f), except that
they depict the semi-magnetic topological insulator and the Chern
insulator phase, respectively. The solid red line is obtained by numerical
calculations. The blue circle and green triangle points are obtained
from the effective Hamiltonians in Eqs.\ (\ref{eq:Effective_model_1})
and (\ref{eq:Effective_model_2}), respectively. The black dashed
lines correspond to the half-quantized value with $\sigma_{1}^{v}\left(k_{r}\right)=\pm e^{2}/2h$.
The red dashed line corresponds to $k_{r}=k_{c}.$ (d, i) and (e,
j) are the same as (a, f), except that they depict the second and
third highest valence bands for the axion insulator phase. The results
for the semi-magnetic topological insulator phase and the Chern insulator
phase are similar to the axion insulator phase. The parameters are
$\Delta_{b}=-0.2$ and $\Delta_{t}=0.05$ for the axion insulator
phase, $\Delta_{t}=0$ for the semi-magnetic topological insulator
phase, and $\Delta_{t}=-0.05$ for the Chern insulator phase. The
film thickness is taken as $n_{z}=10$.}
\end{figure*}

\section{Berry curvature distribution}

Now, we investigate the Berry curvature distributions and search for
the underlying half quantization in the three systems. The first row
in Fig.\ \ref{Fig_BerryCurvature} shows the Berry curvature distribution
$\Omega(k_{x},k_{y})$ as a function of $k_{x}$ and $k_{y}$. The
Berry curvature distributions exhibit distinct behaviors for $k_{r}<k_{c}$
(the low-energy regime) and $k_{r}>k_{c}$ (the high-energy regime),
where $k_{r}$ is the radius of the circle centered at the origin
in the Brillouin zone and $k_{c}$ corresponds to the critical radius
{[}see the red circle in Fig.\ \ref{Fig_BerryCurvature}(b){]}. The
second column shows the conductance of the $m$-th band $\sigma_{m}\left(k_{r}\right)$
as a function of $k_{r}$ at $E_{F}=0$, where
\begin{align}
\sigma_{m}\left(k_{r}\right) & =-i\hbar e^{2}\times\sum_{n\neq m}\intop_{k^{2}<k_{r}^{2}}\frac{d^{2}k}{\left(2\pi\right)^{2}}\frac{\left\langle m\right|v_{x}\left|n\right\rangle \left\langle n\right|v_{y}\left|m\right\rangle }{E_{m}-E_{n}}\nonumber \\
 & \times\frac{f\left(-E_{n}\right)-f\left(-E_{m}\right)}{E_{m}-E_{n}+i\eta},
\end{align}

Let us focus on the low-energy regime (i.e., $k_{r}<k_{c}$). The
highest valence band is characterized by $\sigma_{1}^{v}\left(k_{c}\right)=0.47e^{2}/h$
for the axion insulator phase {[}Fig.\ \ref{Fig_BerryCurvature}(f){]},
$\sigma_{1}^{v}\left(k_{c}\right)=0$ for the semi-magnetic topological
insulator {[}Fig.\ \ref{Fig_BerryCurvature}(g){]}, and $\sigma_{1}^{v}\left(k_{c}\right)=-0.47e^{2}/h$
for the Chern insulator phase {[}Fig.\ \ref{Fig_BerryCurvature}(h){]}.
The second and third highest valence bands are characterized by $\sigma_{2}^{v}\left(k_{c}\right)=-0.38e^{2}/h$
{[}Fig.\ \ref{Fig_BerryCurvature}(i){]} and $\sigma_{3}^{v}\left(k_{c}\right)=0.35e^{2}/h$
{[}Fig.\ \ref{Fig_BerryCurvature}(j){]} for all three phases. In
the low-energy regime, the calculated Hall conductance from the analytical
low-energy effective model in Eq.\ (\ref{eq:Effective_model_2})
{[}the blue circle points in Figs.\ \ref{Fig_BerryCurvature}(f)-\ref{Fig_BerryCurvature}(i){]}
can reproduce the numerically calculated Hall conductance from the
tight-binding models {[}the solid red lines in Figs.\ \ref{Fig_BerryCurvature}(f)-\ref{Fig_BerryCurvature}(i){]}.
This further confirms that, at low-energy regimes, the topologies
of the systems are dominated by the Dirac fermions on the top and
bottom surfaces.

When the whole Brillouin zone is considered, each band in the axion
and Chern insulator phases can only host a quantized Hall conductance,
i.e., $\sigma_{1}^{v}\left(\pi\right)=0$ for the axion insulator
phase {[}Fig.\ \ref{Fig_BerryCurvature}(f){]} and $\sigma_{1}^{v}\left(\pi\right)=e^{2}/h$
for the Chern insulator phase {[}Fig.\ \ref{Fig_BerryCurvature}(h){]}.
The gapless band in the semi-magnetic topological insulator is characterized
by an exact half-quantized Hall conductance with $\sigma_{1}^{v}\left(\pi\right)=e^{2}/2h$
{[}Fig.\ \ref{Fig_BerryCurvature}(g){]}. The second and third highest
valence bands are characterized by $\sigma_{2,3}^{v}\left(\pi\right)=0$
for all three phases {[}Figs.\ \ref{Fig_BerryCurvature}(i) and \ref{Fig_BerryCurvature}(j){]}.

\section{Half-quantized Hall conductance}

In previous studies based on the low-energy effective model, the surface
states are believed to host a half-quantized Hall conductance when
local time-reversal symmetry is broken. However, this is not realizable
in a realistic system where each band can only host an integer-quantized
Hall conductance. The above scenarios explain how this contradiction
is reconciled in a lattice model. In the axion and Chern insulator
phases, the highest valence band is dominated by a nearly half-quantized
Hall conductance from one surface at low energy, and must be compensated
by another nearly half-quantized Hall conductance from the other surface
at high energy. In the semi-magnetic topological insulator phase,
the highest valence band is dominated by a zero Hall conductance from
one surface at low energy, and is compensated by another half-quantized
Hall conductance from the other surface at high energy. This is because
the half-quantized Hall conductance can only originate from the symmetry-protected
gapless highest valence band, and all the rest bands can only contribute
a quantized Hall conductance.

Moreover, due to the coupling between the low-energy and high-energy
states, the exact half quantization can not be observed in the axion
and Chern insulator phases. As shown in Fig.\ \ref{Fig_BerryCurvature}(f),
the contribution to the Hall conductance $\sigma\left(k_{r}\right)$
from the low-energy massive Dirac cone (i.e., the blue circle dots)
increases with the increasing $k_{r}.$ The exact half quantization
$\sigma\left(k_{r}\right)\rightarrow e^{2}/2h$ shall be achieved
only if $k_{r}\rightarrow\infty$. However, the high-energy state
also contributes to $\sigma\left(k_{r}\right)$ when $k_{r}>k_{c}$,
which prohibits the observation of the exact half quantization. Thus,
only a nearly half quantized Hall conductance $\sigma_{1}^{v}\left(k_{c}\right)=0.47e^{2}/h$
is observed at $k_{r}=k_{c}$. In contrast, the exact half quantization
could only be observed in the semi-magnetic topological phase, because
its low-energy state does not contribute to the Hall conductance.

\section{Layer-resolved Hall conductance}

\begin{figure}
\includegraphics[width=8.5cm]{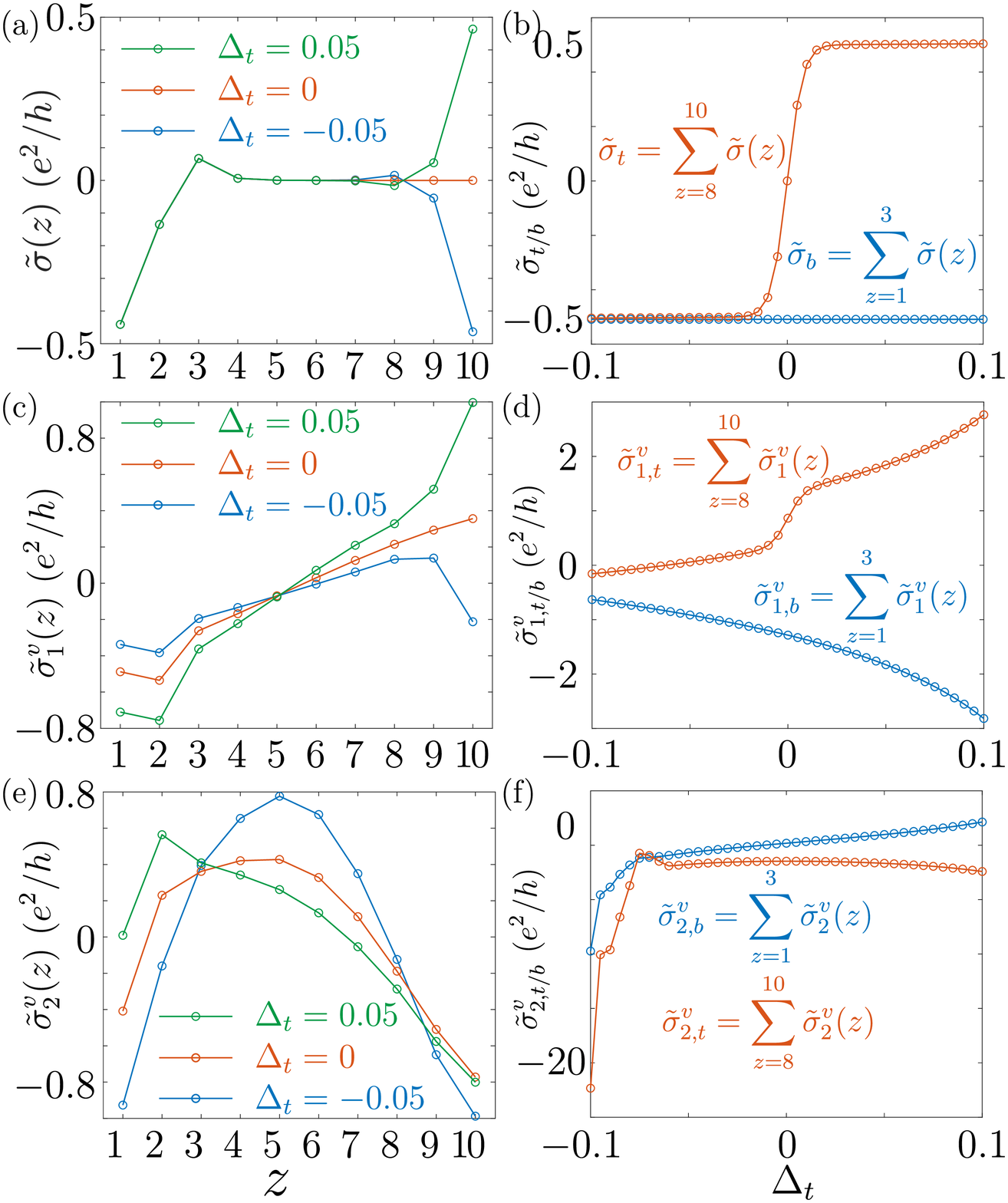}

\caption{\label{Fig_layer_Chern_number}(a) The layer-resolved Hall conductance
$\tilde{\sigma}\left(z\right)$ as a function of layer $z$ for different
$\text{\ensuremath{\Delta_{t}}}$. Green, red, and blue correspond
to the axion insulator ($\text{\ensuremath{\Delta_{t}=0.05}}$), semi-magnetic
insulator ($\ensuremath{\Delta_{t}=0}$), and Chern insulator phases
($\ensuremath{\Delta_{t}=-0.05}$), respectively. (b) $\tilde{\sigma}_{t/b}$
as functions of $\Delta_{t}$. Here, $\tilde{\sigma}_{t/b}$ depicts
the layer-resolved Hall conductance contributed from the top/bottom
surface layers. In (a-b), the calculations take into account the contributions
from all the occupied bands. (c-d) and (e-f) are the same as (a-b),
except that the obtained layer-resolved Hall conductance only takes
into account the contributions from the first valence band in (c-d)
and from the second valence band in (e-f), respectively. Here, we
take the Fermi energy $E_{F}=0$, the film thickness $n_{z}=10$,
and $\Delta_{b}=-0.2$.}
\end{figure}

The axion insulator phase can be distinguished from the normal insulator
phase by calculating the layer-resolved Hall conductance. The layer-resolved
Hall conductance of the $m$-th occupied band is given by\,\citep{FuB2021PRR,Wang15prbrc,Mong10prb,Varnava18prb,Essin09prl}
\begin{equation}
\tilde{\sigma}_{m}\left(z\right)=\frac{e^{2}}{2\pi h}\int d^{2}\mathbf{k}\mathcal{F}_{xy}^{mm}\left(\mathbf{k},z\right),
\end{equation}
where
\begin{equation}
\mathcal{F}_{\alpha\beta}^{mn}\left(\mathbf{k}\right)=\partial_{\alpha}\mathcal{A}_{\beta}^{mn}\left(\mathbf{k}\right)-\partial_{\beta}\mathcal{A}_{\alpha}^{mn}\left(\mathbf{k}\right)+i\left[\mathcal{A}_{\alpha}^{mn}\left(\mathbf{\mathbf{k}}\right),\mathcal{A}_{\beta}^{mn}\left(\mathbf{\mathbf{k}}\right)\right]
\end{equation}
is the non-Abelian Berry curvature in terms of $\mathcal{A}_{\alpha}^{mn}\left(\mathbf{\mathbf{k}},z\right)$
with the band indexes $m$ and $n$. The layer-resolved Hall conductance
of all occupied bands is given by $\tilde{\sigma}\left(z\right)=\sum_{E_{m}<E_{F}}\tilde{\sigma}_{m}\left(z\right)$.
Moreover, the net Hall conductance of the $m$-th band and the layer-resolved
Hall conductance of the $m$-th band is connected through $\sigma_{m}=\sum_{z=1}^{n_{z}}\tilde{\sigma}_{m}\left(z\right)$.

Figure\ \ref{Fig_layer_Chern_number}(a) shows $\tilde{\sigma}\left(z\right)$
as a function of $z$ for the axion insulator ($\Delta_{t}=0.05$),
semi-magnetic topological insulator ($\Delta_{t}=0$), and Chern insulator
phases ($\Delta_{t}=-0.05$). The bottom surface layers have the same
magnetization alignments in the three phases, thus they host the same
layer-resolved Hall conductance {[}$\tilde{\sigma}\left(z=1,2,3\right)$
in Fig.\ \ref{Fig_layer_Chern_number}(a){]}. The top surface layers
have the opposite magnetization alignments in the axion insulator
(green) and Chern insulator (blue) phases, thus the corresponding
layer-resolved Hall conductances have the opposite signs in the two
phases {[}$\tilde{\sigma}\left(z=8,9,10\right)$ in Fig.\ \ref{Fig_layer_Chern_number}(a){]}.
The top surface layers are non-mangnetic in the semi-magnetic topological
insulator (red) phase, thus the corresponding layer-resolved Hall
conductances is zero. This phenomenon is observed more clearly in
Fig.\ \ref{Fig_layer_Chern_number}(b), where we show $\tilde{\sigma}_{t/b}$
as functions of $\Delta_{t}$. $\tilde{\sigma}_{t}=\sum_{z=8}^{10}\tilde{\sigma}\left(z\right)$
and $\tilde{\sigma}_{b}=\sum_{z=1}^{3}\tilde{\sigma}\left(z\right)$
depict the layer-resolved Hall conductance of the top and bottom surface
layers, respectively. With the increasing $\Delta_{t}$, $\tilde{\sigma}_{b}$
remains unchanged but $\tilde{\sigma}_{t}$ increases from $-0.5$
to $0$ and then to $0.5$, which corresponds to the phase crossovers
from the Chern insulator phase to the semi-magnetic topological insulator
phase and then to the axion insulator phase.

The above calculations take into account the contributions from all
the occupied bands and we show that the half quantization can be extracted
in the three phases. However, we find that the half quantization cannot
be extracted if only one single band is considered. This is shown
in Figs.\ \ref{Fig_layer_Chern_number}(c-d) and Figs.\ \ref{Fig_layer_Chern_number}(e-f),
which are the same as Figs.\ \ref{Fig_layer_Chern_number}(a-b),
expect that the numerically calculated layer-resolved Hall conductance
only considers the contributions from the first valence band and the
second valence band, respectively. The layer-resolved Hall conductances
of one single band have divergent values for each layer. This implies
that the half-quantized surface Hall effect is a consequence of all
the occupied bands, rather than the individual gapped surface bands.

\section{Different thickness and parameters}

\begin{figure}
\includegraphics[width=8.5cm]{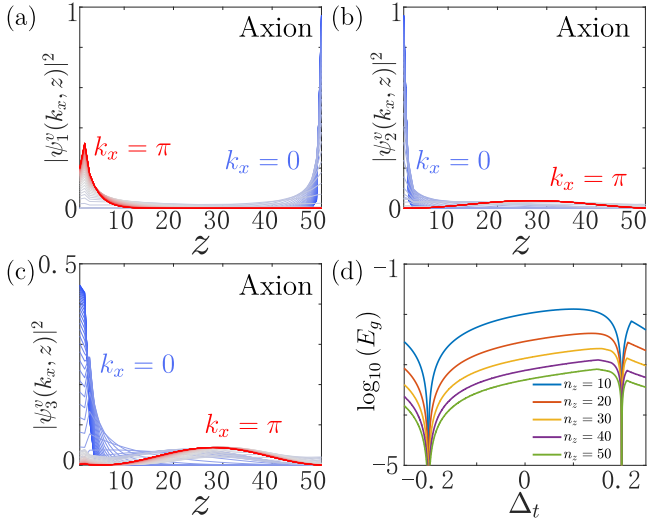}

\caption{\label{Fig_thickness}(a) Wave function distribution as a function
of layer index $z$ for the highest valence band of the axion insulator
phase for different $k_{x}.$ (b) and (c) are the same as (a), except
that they depict the second and third valence bands. The wave function
distributions of the semi-magnetic topological insulator phase and
the Chern insulator phase are similar to the axion insulator phase.
(d) The logarithm of the energy gap between the first and second valence
bands as a function of $\Delta_{t}$ for different thicknesses $n_{z}$.
Each point is obtained by find the minimum energy difference between
the first and second valence bands by scanning in the two-dimensional
Brillouin zone. The $k$ points used in the calculation is $N_{k}\times N_{k},$with
$N_{k}=200$. The parameters are taken as $\Delta_{b}=-0.2$. In (a-c),
we take $\Delta_{t}=0.05$ and $n_{z}=50$.}
\end{figure}

\begin{figure}[t]
\includegraphics[width=8.5cm]{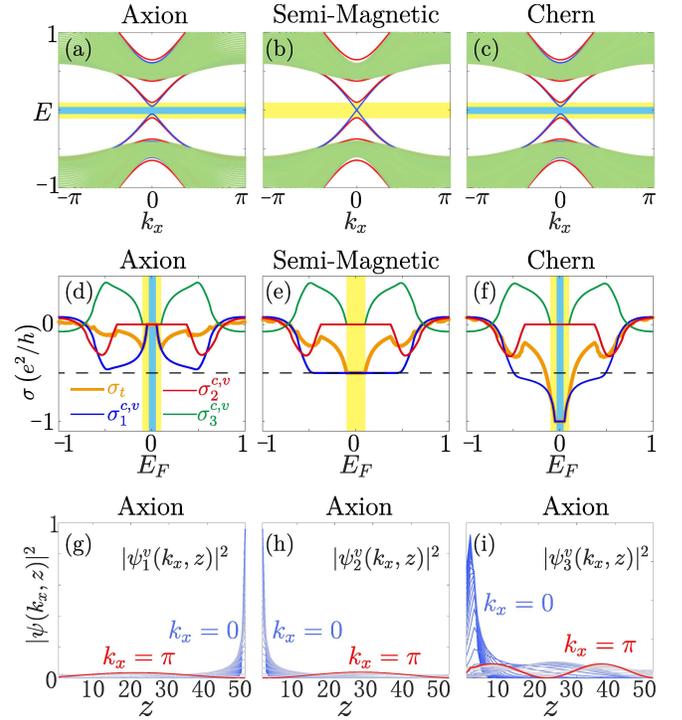}

\caption{\label{Fig_Parameter1}(a) Energy spectra of the axion insulator phase.
Here the color scheme of the bands indicates the wave function distribution.
(d) Numerically calculated Hall conductance $\sigma_{1}^{c,v}$ (blue),
$\sigma_{2}^{c,v}$ (red), $\sigma_{3}^{c,v}$ (green), and $\sigma_{t}$
(orange) as a function of $E_{F}$. Here, $\sigma_{t}$ depicts the
conductance contributed from all the bands, $\sigma_{1}^{c,v}$ depict
the conductance contributed from the lowest conduction and highest
valence bands, $\sigma_{2}^{c,v}$ depict the conductance contributed
from the second lowest conduction and second highest valence bands,
and so on. The solid black line corresponds to $\sigma=e^{2}/2h.$
In (a) and (d), the yellow and light blue regions correspond to the
magnetic gap on the top and bottom surfaces, respectively. (b, e)
and (c, f) are the same as (a, d), except that they depict the semi-magnetic
topological insulator phase and the Chern insulator phase, respectively.
(g) Probability distribution as a function of layer index $z$ for
the highest valence band of the axion insulator phase for different
$k_{x}.$ (h) and (i) are the same as (a), except that they depict
the second and third valence bands. The wave function distributions
of the semi-magnetic topological insulator phase and the Chern insulator
phase are similar to the axion insulator phase. We take $\Delta_{t}=0.05$
for the axion insulator phase, $\Delta_{t}=0$ for the semi-magnetic
topological insulator phase, and $\Delta_{t}=-0.05$ for the Chern
insulator phase. The Zeeman splitting term for the bottom surface
is $\Delta_{b}=-0.1$ for all three phases. The film thickness is
taken as $n_{z}=50$.}
\end{figure}

Above, we consider a thinfilm case with $n_{z}=10$. Now, we show
that the conclusions are irrelevant to the film thickness. Figures\ \ref{Fig_thickness}(a-c)
show the probability distributions of the first, second, and third
highest valence bands for different $k_{z}$ with $n_{z}=50$, respectively.
The conclusions are similar to the thinfilm case shown in Figs.\ \ref{Fig_wavefunction}(a-c).
Figure\ \ref{Fig_thickness}(d) shows the logarithm of the energy
gap between the first and second valence bands as a function of $\Delta_{t}$
for different thicknesses $n_{z}$. Though the energy difference decreases
with the increasing film thickness, the first and second valence bands
are still well separated in energy scale, as long as $\Delta_{t}\neq\pm\Delta_{b}$.

Moreover, we consider the case for $\Delta_{b}=-0.1$. The first row
in Fig.\ \ref{Fig_Parameter1} shows the spectra of the three topological
phases, which are similar to those shown in the second row in Fig.\ \ref{Fig_illustration},
except that the high energies for the first and second highest valence
and lowest conduction bands merge into the bulk. This can be observed
more clearly in the third row in Fig.\ \ref{Fig_Parameter1}, which
shows the probability distributions of the first, second, and third
highest valence bands, respectively. Thus, the probability distribution
of the highest valence band at high energy with $k_{x}=\pi$ can be
localized not only at the surface {[}the red lines in Figs.\ \ref{Fig_wavefunction}(a)
and \ref{Fig_thickness}(a){]}, but also at the bulk {[}the red line
in Fig.\ \ref{Fig_Parameter1}(g){]}. Moreover, by checking the Hall
conductance (the second row in Fig.\ \ref{Fig_Parameter1}), the
probability distributions of the state at high energy will not change
the fact that the Hall conductance is dominated by the lowest conduction
and highest valence bands, when the Fermi energy resides inside the
magnetic gap.

\section{Conclusion}

In this work, we have investigated the energy spectra, wave function
distribution, and the corresponding Berry curvature in the three distinct
topological phases in magnetic topological insulator films, including
the axion insulator phase, the semi-magnetic topological insulator
phase, and the Chern insulator phase. In the axion and Chern insulator
phases, a nearly half-quantized Hall conductance is observed at low
energy near $\Gamma$ point for the gapped Dirac cone (which is in
accordance with the previous consensus), however, another nearly half-quantized
Hall conductance must be compensated at high energy away from $\Gamma$
point to ensure that the total Hall conductance is an integer-quantized
number in units of $e^{2}/h$. In the semi-magnetic topological insulator
phase, the gapless Dirac cone hosts a vanishing Hall conductance at
low energy near $\Gamma$ point as expected, however, another half-quantized
Hall conductance emerges at high energy away from $\Gamma$ point.
This explains how the contradiction between the previous consensus
based on the low-energy effective model and the scenarios based on
a lattice model is reconciled. Moreover, due to the coupling between
the low-energy state and the high-energy state, the exact half quantization
is only revealed in the gapless surface state in the semi-magnetic topological
insulator phase, but not found in the gapped surface state in the
axion and Chern insulator phases from the band calculation.

On the other hand, we adopt the layer-resolved Hall conductance to
characterize the three phases. The half-quantized surface Hall effect
is revealed in the three topological phases and it is contributed
by all the occupied bands. This is distinct to the previous consensus
that the half quantization originates from the individual gapped band.
\begin{acknowledgments}
We thank helpful discussions with Bo Fu and Huan-Wen Wang. This work
was supported by the Research Grants Council, University Grants Committee,
Hong Kong under Grant Nos. C7012-21G and 17301220 and the National
Key R\&D Program of China under Grant No. 2019YFA0308603.
\end{acknowledgments}

\bibliographystyle{apsrev4-1-etal-title_6authors}
\bibliography{refs-transport}

\end{document}